# Applying convex layers, nearest neighbor and triangle inequality to the Traveling Salesman Problem (TSP)

Liew Sing
liews_ryan@yahoo.com.sg

March 27, 2012

**Abstract**

The author would like to propose a simple but yet effective method, convex layers, nearest neighbor and triangle inequality, to approach the Traveling Salesman Problem (TSP). No computer is needed in this method. This method is designed for plain folks who faced the TSP everyday but do not have the sophisticated knowledge of computer science, programming language or applied mathematics. The author also hopes that it would give some insights to researchers who are interested in the TSP.

### 1.Introduction

The Traveling Salesman Problem (TSP) is stated as the follows. A salesman is required to visit each of *n* given cities once and only once, starting from any city and returning to the original city of departure. The problem is how should he travel in order to minimize the total travel distance[7] assuming that the shortest traveling distance (optimal tour) would keep the traveling cost, e.g. fuel fees if the salesman travels by car, to the minimum? The TSP is a very common problem among all of us and we may face it everyday without noticing it. For instance, a student would like to have a visit of all the faculties of a university and return to the first faculty where he departed. How should the student plan the routes in order to have an optimal tour? Unfortunately, the TSP in general is very hard to solve and it is one of the most widely studied NP-hard combinatorial optimization problem[10]. The difficulty becomes obvious when one considers the number of possible tours by the method of brute force searching even for a relatively small number of cities *n*. For instance, for a problem with 20 cities (*n*=20) by brute force searching, it would be (20-1)!/2 tours, which is more than $10^{18}$ tours! The TSP is a class of difficult problems whose time complexity is widely believed exponential. Despite the fact that the TSP is a very common problem and may happen to any one of us when we have to travel to various places, ironically up to date, there are only two classes of algorithms or so-called solutions in solving the TSP: *Exact* algorithms and *Approximate* (or *heuristic*) algorithms[7] which required sophisticated knowledge of computer science, programming language or applied mathematics e.g. linear programming. In some cases, like cutting-plane method from *Exact* algorithms, the average time to work on a nine cities problem by hand of a scholar in 1955 was a about three hours[11].

What if plain folks like sales executive or lawyer who need to travel to various places to meet clients know nothing about computer science, programming language or linear programming? It took a scholar three hours to look for an optimal tour of nine cities, let alone a normal folk. Thus, the author would like to address this concern in this paper and therefore propose a method which no computer is needed and only require the tour seeker a pencil, an eraser and the understanding of convex layers, nearest neighbor and triangle inequality.

The sequences of this paper are as follows. Firstly, the author introduces the construction of convex layers. Then, the author introduces the concepts of nearest neighbor as well as triangle inequality. Thirdly, the author demonstrates the ideas on the classic 48-States-Problem and lastly, the paper ends with conclusions and discussions.

## 2.Convex layers
Before the author proceeds to discuss convex layers, the author would like to draw the reader's attention on convex hull first. Convex layers are constructed from convex hull, after all. In mathematical sense, we can descript a convex hull as the smallest convex polygon that encloses all the vertices (or points) in a set[12] or, in simple words, we can imagine convex hull as how a rubber band will enclose a set of points[2], as shown in Figure 1. In Figure 1, it illustrates a convex hull that enclosed a set of square points. In general, we called these square points as interior points. Notice that we can actually construct another convex hull for the interior points, that is, we have constructed convex layers as illustrated in Figure 2. In computational geometry, the algorithm of constructing convex layers is called Onion-Peeling[8] and it is stated as follows[3]. Consider a set $S$ of $n$ on a 2-D plane. Compute the convex hull of $S$, and let $S'$ be the set of points remaining in the interior of the hull. Then compute the convex hull of $S'$ and recursively repeat this process until no more points remain. Readers who are interested in the details of the Onion-Peeling may wish to study [3].

## 3.Nearest neighbor
Nearest neighbor is a term that adopted from "nearest neighbor search" which were originated from computer programming but the idea is relatively very simple. In words, the idea can be explained as follows. In nearest neighbor search, the salesman that required to travel various cities (or points or vertices) would have to choose any city to start with and drive to the closest city among those not yet visited[2]. Figure 3 illustrates the directions (start from the red dot) of the first few points that a salesman would travel based on the method of nearest neighbor[2].

## 4.Triangle inequality
In mathematical sense, the triangle inequality states that for any triangle, the sum of any two edges (or lengths) must be greater than or equal to the remaining edge. That is, the distance travel from city A to city B $d(A,B)$ plus the distance travel from city B to city C $d(B,C)$ must not be less than the distance travel from city A to city C $d(A,C)$ directly[2]. Figure 4 illustrates the basic principle of triangle inequality: $d(A,B)+ d(B,C) \geq d(A,C)$[1].

## 5.Demonstration with the classic 48-States-Problem.
The difficulty of the TSP was in fact noticed in 1930 by Karl Menger, an Austrian mathematician and economist, who first brought the challenge of the TSP to the mathematics community[2]. The author would like to divert the attention of history or everything about the TSP to the monograph of [1]. In 1954, Dantzig, Fulkerson and Johnson published a paper to demonstrate a method to solve the 48-States-Problem[4]

(Figure 5). The methods used were *Exact* algorithms which based on linear programming and cutting-plane method[2]. In general, *Exact* algorithms are guaranteed to find the optimal solution in a bounded number of steps but unfortunately also complex with codes in programming language and very demanding of computer power[7]. Readers who are interested in the details of the methods may wish to study Dantzig *et al*'s [4] or the monograph of [1].

Now, let's us demonstrate the method the author proposed. There are three major principles, namely convex layers, nearest neighbor and triangle inequality, in this proposed method and it works as follows. In step one, we construct the convex layers for the problem given. In step two, we merge the convex layers by nearest neighbor. We repeat step two until a hamiltonian tour or "product" to be found. Note that the product obtained, in general, is not the optimal tour. However, if the problem given has only two convex layers, the product obtained will be very close to the optimal tour. In special cases, we may obtain the optimal tour immediately. Lastly, we use triangle inequality as a tool to assist us to spot the area that required tour improvement on the product (step three).

Figure 6 illustrates the convex layers for the classic 48-States-Problem. Notice that there are six convex layers in the problem. In addition, we named the outermost layer as $1^{st}$ layer, the second outermost layer as $2^{nd}$ layer and so on. The innermost layer will be $6^{th}$ layer. Next, we proceed to step two to start merging the $1^{st}$ and $2^{nd}$ layer based on the principle of nearest neighbor. Recall that, by definition of the nearest neighbor, the salesman that required to travel various cities (or points or vertices) would have to choose any city to start with and drive to the closest city among those not yet visited[2]. So, in merging convex layers, we start from randomly choose any vertex form the $1^{st}$ layer, erase the longer edge (every vertex has two edges) and connect the vertex to the nearest $2^{nd}$ layer's vertex. Figure 7 illustrates the product of the $1^{st}$ and $2^{nd}$ layer. Next, we repeat step two until the hamiltonian path to be found. Figure 8, Figure 9, Figure 10 and Figure 11 illustrate the merging of the convex layers until the hamiltonian path was found. After the hamiltonian path was found, we proceed to step three, where we apply triangle inequality to spot the area that require tour improvement of the hamiltonian path. Notice that in Figure 11, there is a red circle that indicate the area that require tour improvement. For instance, based on triangle inequality, it is obvious that the edges that connect Raleigh, Charleston and Richmond will require improvement. As a matter of fact, tour improvement is a very tricky business without the help of a computer to calculate the distance and it may require the tour seeker to ponder for the solution. Triangle inequality merely serves us as a tool to help us to spot the potential area that requires tour improvement. In *Approximate* (or *heuristic*) algorithms, Lin-Kernighan algorithms are particularly useful in our case study here and it is considered as one of the most effective tour improvement algorithms. Reader who are interested in Lin-Kernighan algorithms may seek [2] or [9] for details. Figure 12 illustrates the optimal tour of the 48-States-Problem presented by Dantzig *et al* on 1954[4].

**6. Discussions and conclusions**
Recall the definitions of convex hull: convex hull as the smallest convex polygon that encloses all the vertices (or points) in a set[12]. In other words, if the TSP has only one convex layer, convex hull will be the optimal (shortest) tour[6]! Convex hull is indeed a very good idea for the TSP, however when the TSP has two convex layers and get merged by the nearest neighbor, we observed that in general the error of the hamiltonian path would start to increase. So this observation gives rise to another question: If we have *n* layers of TSP, by merging the layers in the same fashion, will the error of the hamiltonian path increase in linear ($Cn$) or exponential ($C^n$) (for *C* is a constant)? If it increases in linear, it will be great because we can reduce the error by computer. If it increases in exponential, it will be bad. That is, the method propose here would be a bad idea or there is still a lot of room for improvement in merging convex layers with nearest neighbor. However, based on the demonstration given here, we believe that the error would increase in the fashion of linear instead of exponential. Researchers who are interested in merging convex layers may want to look into merging convex layers with other alternatives.

Despite the fact that merging convex layers with nearest neighbor does not provide an immediate optimal solution, it provides a simple and effective method for the plain folks who are seeking an optimal tour. From the table of "Road Distances Between Cities In Adjusted Units" published by Dantzig *et al* on 1954[4], the optimal tour is 699 units. A "pure" nearest neighbor method will gives 1,013 units[2], a greedy tour will gives 995 units[2], a farther-insertion tour will gives 778 units and a Christofides tour will gives 759 units[2]. So it is obvious that the method we proposed here (without tour improvement step three) is better than the "pure" nearest neighbor method, greedy tour, farther-insertion tour as well as Christofides tours. The only drawback of the method, as mentioned earlier, it requires the tour seeker to spot for the area that needs tour improvement and the tour seeker has to ponder for a solution. However, we observe that we may use "area" as another tool for tour improvement. In Figure 13, we can see that the tour on the left is better than the right, but at the same time, the area of the tour on the left is also bigger than the one on the right. By the same argument, we recommend the tour seeker keeps area in mind and try to improve the tour by having as bigger area as possible. Notice that the area of the optimal tour from Dantzig *et al* is bigger than the area of the tour we proposed here. In addition, tour seeker should also prevent any cross edges as it will render a nonoptimal tour[5].

In conclusion, the author proposed a simple but yet effective method to the general plain folks who are not sophisticated in computer science, programming language or linear programming but yet face the TSP very often. The method proposed here does not require any computer or any sophisticated knowledge but just three simple principles: convex layers, nearest neighbor and triangle inequality. On the other hand, there is a drawback in this method that the tour seeker has to ponder a solution him/herself for the tour improvement. Nevertheless, the author has also suggested some "guidelines" of using area and edges-crossing prevention to help the tour seeker to improve the tour. Last but not least, the author has also raises the concern of having this proposal to implement in computer science and the author believed that it would be helpful to the community who are interested in the TSP.

**7.Reference**

## 8.Appendix
Below are the figures for Applying convex layers, nearest neighbor and triangle inequality to the Traveling Salesman Problem (TSP)

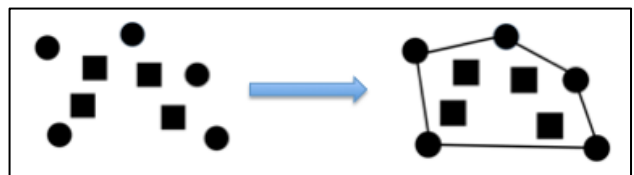
Figure 1: A convex hull with four interior points (square).

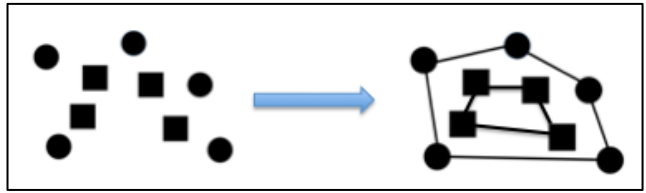

Figure 2: Convex layers of nine points.

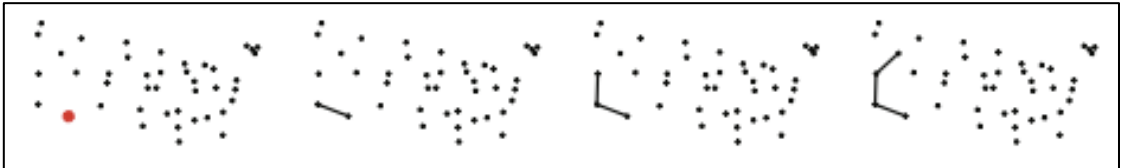

Figure 3: Nearest Neighbor search.

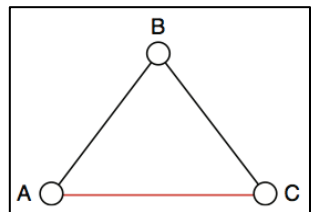

Figure 4: Triangle Inequality.

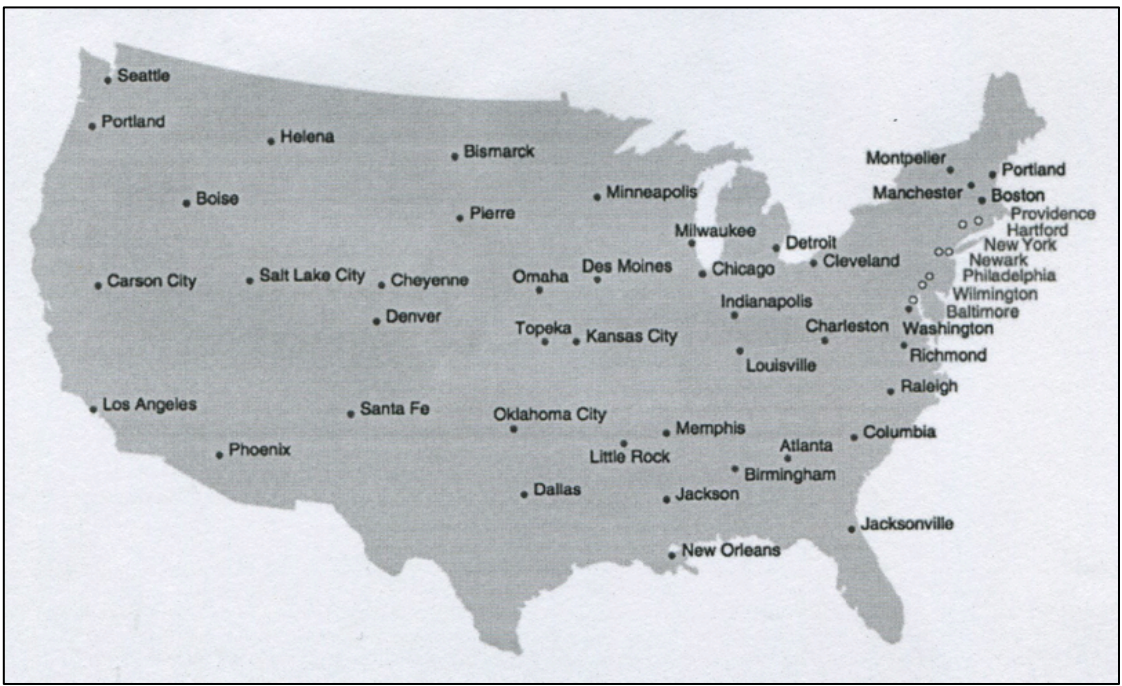

Figure 5: The classic 48-States-Problem.

Figure 6: The convex layers of the classic 48-States-Problem.

Figure 7: The outermost layer is the product of the 1st and 2nd layer. The inner layer is the original 3rd layer.

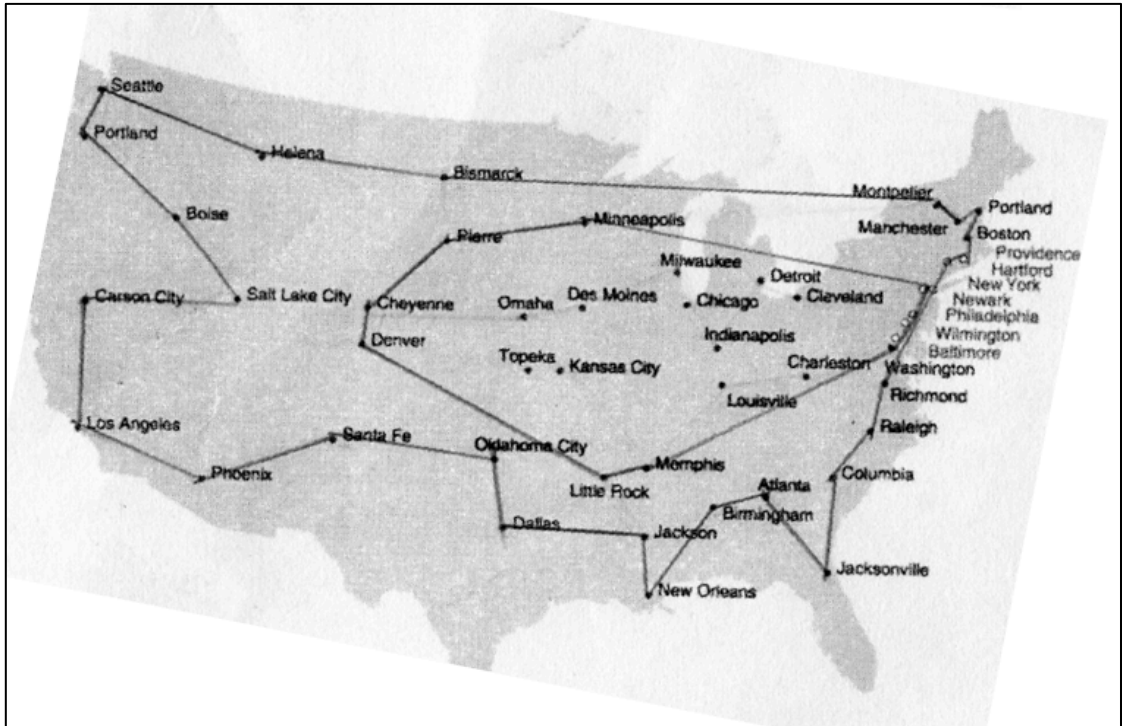

Figure 8: The outermost layer is the product of 1$^{st}$, 2$^{nd}$ and 3$^{rd}$ layer. The innermost layer is the original 4$^{th}$ layer.

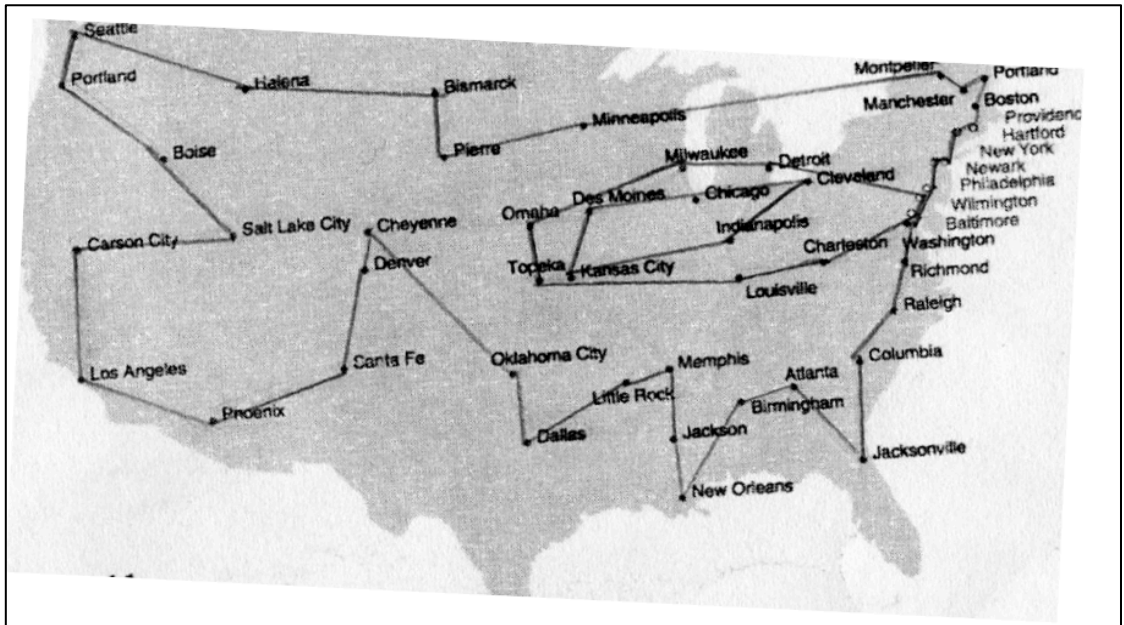

Figure 9: The outermost layer is the product of the 1$^{st}$, 2$^{nd}$, 3$^{rd}$ and 4$^{th}$ layer. The second outermost layer is the original 5$^{th}$ layer. The innermost layer is the original 6$^{th}$ layer.

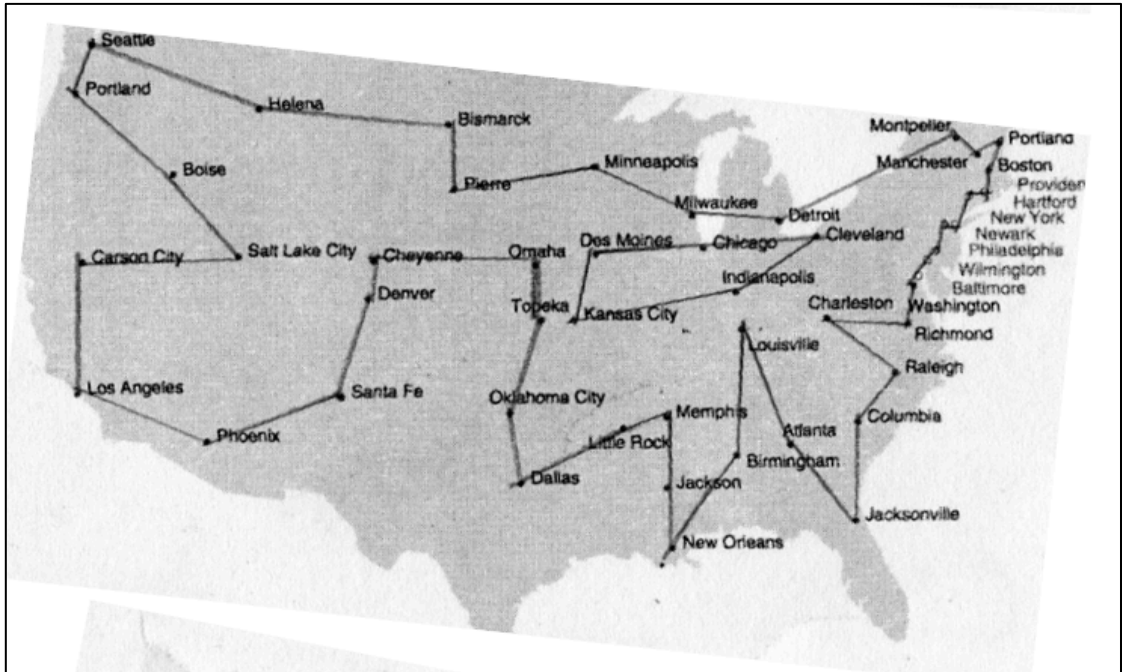

Figure 10: The outermost layer is the product of the 1$^{st}$, 2$^{nd}$, 3$^{rd}$, 4$^{th}$ and 5$^{th}$ layer. The innermost layer is the original 6$^{th}$ layer.

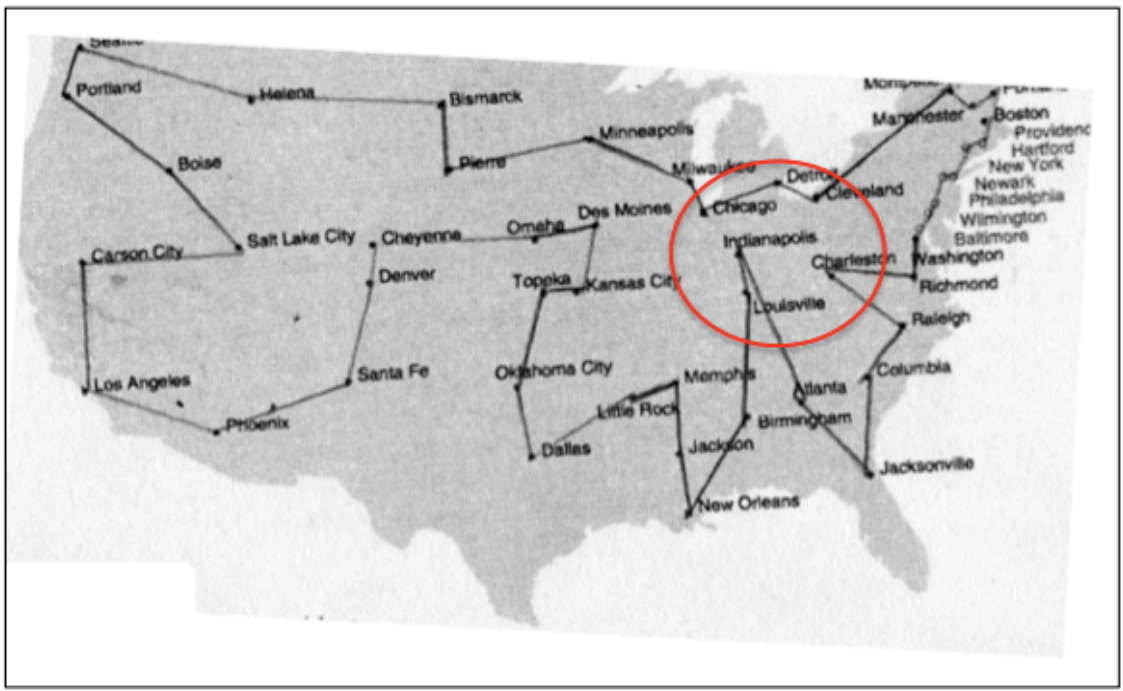

Figure 11: The final product of the convex layers. The red cycle spotted the area that required step three triangle inequality for tour improvement.

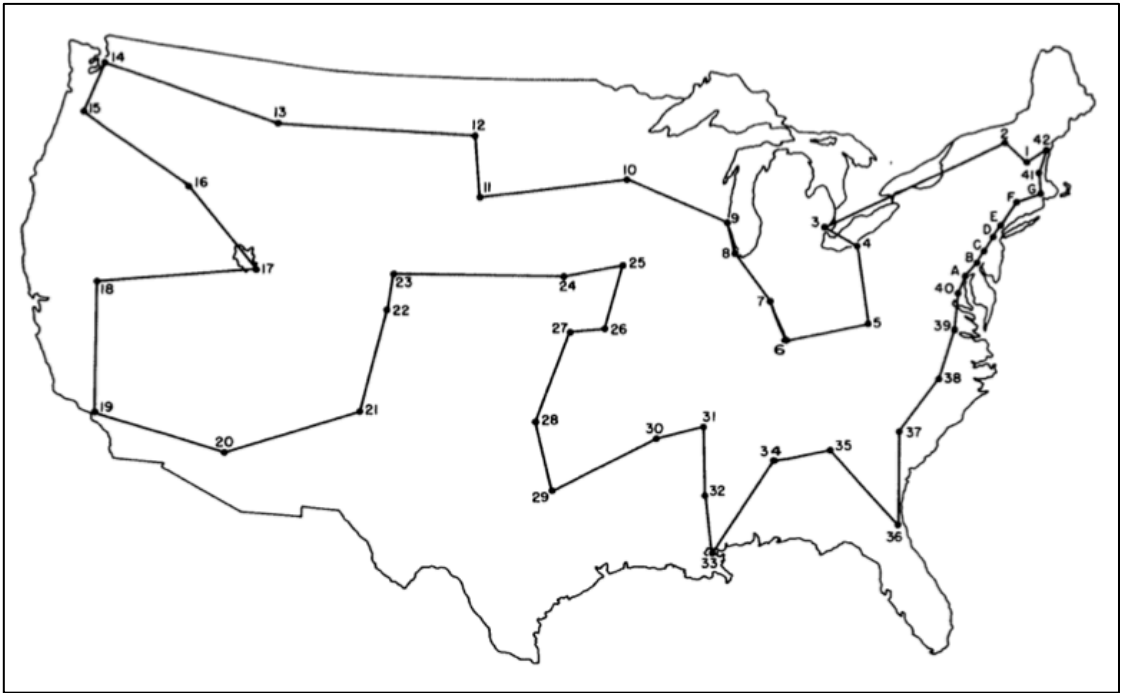
Figure 12: The optimal solution from Dantzig, Fulkerson and Johnson (1954).

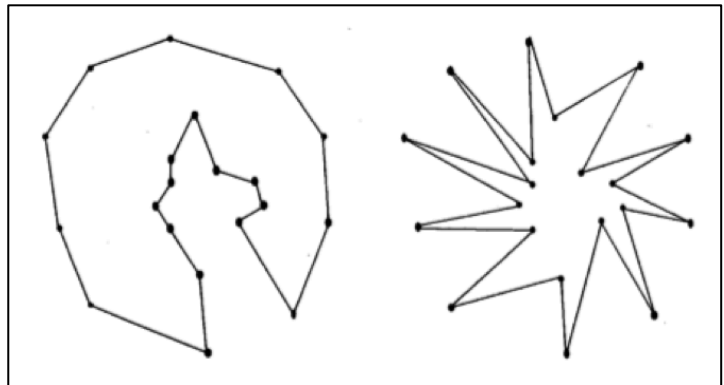
Figure 13: Better tour give raise to bigger area.